\documentclass[%
 reprint,
 amsmath,amssymb,
 aps,
]{revtex4-2}

\usepackage{graphicx}
\usepackage{dcolumn}
\usepackage{bm}

\usepackage{graphicx}
\usepackage{epstopdf, epsfig}
\usepackage{lipsum}
\usepackage{amsmath,amssymb}
\usepackage{amsfonts}
\usepackage{graphicx}
\usepackage{todonotes}
\usepackage{tikz}
\usepackage{color}
\usetikzlibrary{shapes}

\definecolor{MATLABcyan}{RGB}{77, 190, 238}
\definecolor{MATLABred}{RGB}{217, 83, 25}
\definecolor{MATLABgreen}{RGB}{119, 172, 48}
\definecolor{MATLAByellow}{RGB}{237, 177, 32}
\definecolor{MATLABpurple}{RGB}{126, 47, 142}
\definecolor{MATLABblue}{RGB}{0, 114, 189}
\definecolor{MATLABgrey}{RGB}{128, 128, 128}

\begin{document}

\preprint{APS/123-QED}






\title[]{Asymmetric limit cycles within Lorenz chaos induce anomalous mobility for a memory-driven active particle}

\author{Rahil N. Valani$^{1}$}\email{rahil.valani@physics.ox.ac.uk}
\author{Bruno S. Dandogbessi$^{2}$}
\affiliation{$^1$Rudolf Peierls Centre for Theoretical Physics, Parks Road, University of Oxford, OX1 3PU, United Kingdom}
\affiliation{$^2$Ecole Nationale Supérieure de Génie Mathématique et Modélisation (ENSGMM), Université Nationale des Sciences, Technologies, Ingénierie et Mathématiques, Abomey, Republique du Bénin}

\date{\today}

\begin{abstract}

On applying a small bias force, non-equilibrium systems may respond in paradoxical ways such as with giant negative mobility (GNM) -- a large net drift opposite to the applied bias, or giant positive mobility (GPM) -- an anomalously large drift in the same direction as the applied bias. Such behaviors have been extensively studied in idealized models of externally driven passive inertial particles. Here, we consider a minimal model of a memory-driven active particle inspired from experiments with walking and superwalking droplets, whose equation of motion maps to the celebrated Lorenz system. By adding a small bias force to this Lorenz model for the active particle, we uncover a dynamical mechanism for simultaneous emergence of GNM and GPM in the parameter space. Within the chaotic sea of the parameter space, a symmetric pair of coexisting asymmetric limit cycles separate and migrate under applied bias force, resulting in anomalous transport behaviors that are sensitive to the active particle's memory. Our work highlights a general dynamical mechanism for the emergence of anomalous transport behaviors for active particles described by low-dimensional nonlinear models.

\end{abstract}

\maketitle

\textit{Introduction.} When a small bias force is applied to a system at equilibrium, it responds with a net drift in the direction of the applied bias. However, non-equilibrium systems can respond in counterintuitive ways to a small applied bias. A peculiar anomalous transport behavior that defies intuition is negative mobility, where a system with no net drift in the absence of a bias force, responds with a net drift in the direction opposite to an applied small bias force. Negative mobility has been observed in experiments of both classical as well as quantum systems, such as electrons in GaAs quantum wells~\citep{PhysRevLett.56.2736}, transport through semiconductor superlattices~\citep{PhysRevLett.75.4102,PhysRevLett.85.1302} and particles moving in structured microfluidic channels~\citep{Ros2005}. Theoretically and numerically, negative mobility and related anomalous transport behaviors have been extensively investigated in various minimal models of a passive inertial particle, either deterministic or stochastic, with typical features of dissipation, spatially periodic potential and external time-periodic driving~\citep{PhysRevLett.88.190601,PhysRevE.66.066132,PhysRevE.92.062903,PhysRevLett.98.040601,Spiechowicz2024,PhysRevLett.98.040601,memoryANM,FANG2022111775,Guo_2017,Wiśniewski_2022,ANMnonchaos1,ANMnonchaos2,10.1063/5.0146649,Spiechowicz_2019}. These studies have identified different mechanisms that give rise to negative mobility in these externally-driven non-autonomous systems, and the effect is attributed to at least one of the following ingredients: noise-induced fluctuations, chaotic attractors and periodic limit cycle attractors. Investigation of negative mobility in self-driven active systems has been relatively scarce but it is increasing with a recent surge of interest in active matter~\citep{activeNM1,activeNM2,activeANM3,activeANM4,activeANM5,Valani2022ANM}. 

In this letter, we consider a simple Lorenz system model of a memory-driven active particle inspired from the experimental system of walking/superwalking droplets~\citep{Couder2005WalkingDroplets,superwalker}. Although the existence of absolute negative mobility in this model has been reported in an earlier work by the first author~\citep{Valani2022ANM}, here we rationalize its emergence by identifying a general dynamical mechanism. We uncover a dynamical mechanism for simultaneous emergence of giant negative mobility (GNM) and giant positive mobility (GPM) rooted in asymmetric limit cycles of the Lorenz system~\citep{Lorenz1963}.  

\begin{figure}
 \centering
 \includegraphics[width=0.8\columnwidth]{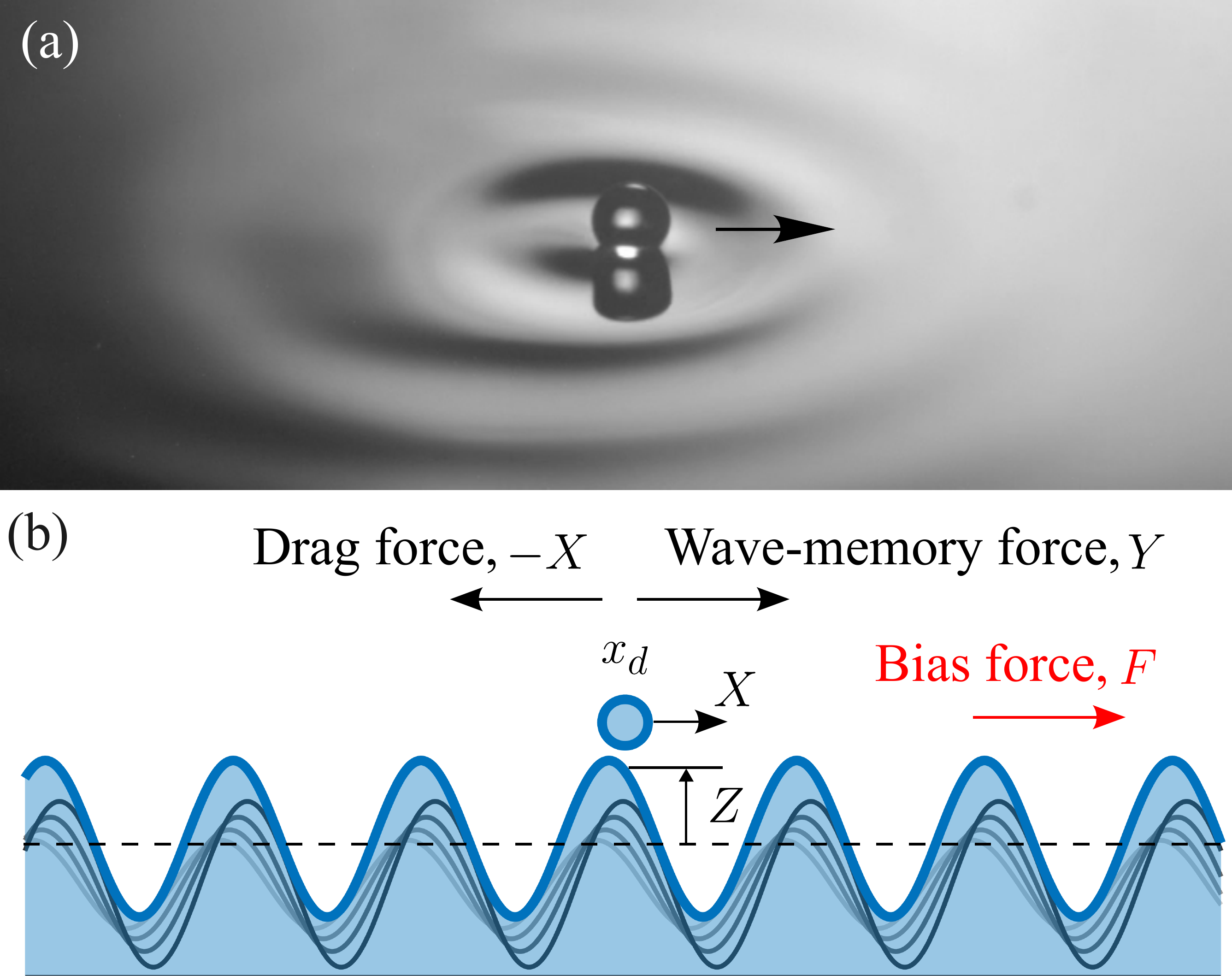}
\caption{Lorenz system model for a memory-driven active particle. (a) Experimental image of a millimeter-sized superwalking droplet~\citep{superwalker} and (b) a schematic (not to scale) of the corresponding memory-driven active particle model. A particle located at $x_d$ (blue circle) moves in one-dimension with horizontal velocity $\dot{x}_d=X$ and experiences a memory force $Y$ from its self-generated wave field (blue) built-up by superposition of individual cosine decaying waves generated by the particle at each instant of time (black and gray curves, with the higher intensity of the color indicating the waves created more recently). In addition, the particle experiences an effective drag force $-X$ and an external constant applied bias force $F$. The wave field height at particle location is $Z$.}
 \label{Fig: schematic}
 \end{figure}

\begin{figure}
 \centering
 \includegraphics[width=\columnwidth]{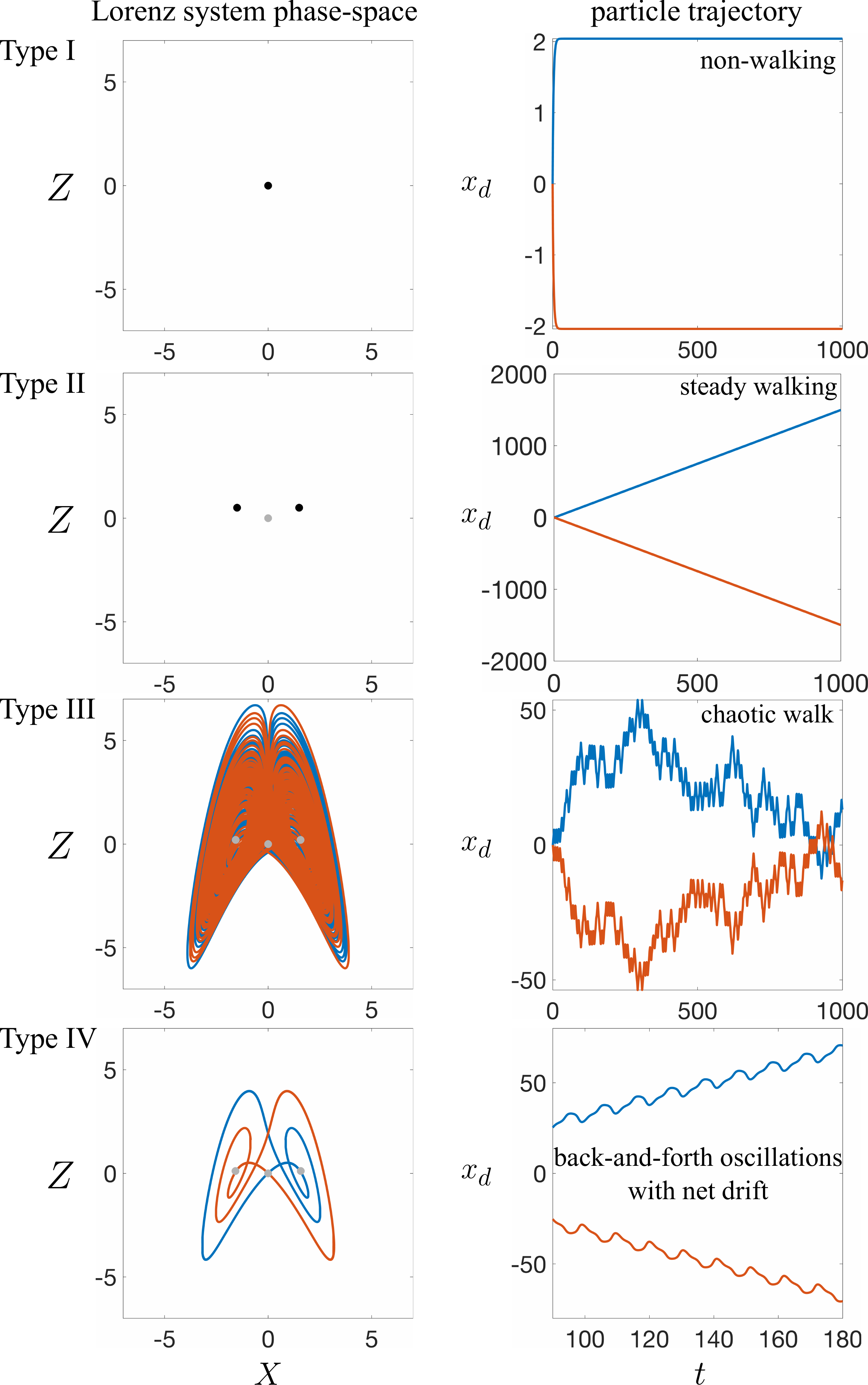}
\caption{Correspondence between phase-space behaviors of the Lorenz system and the trajectories of the memory-driven active particle. Phase-plane projection in the $(X,Z)$ plane (left) and space-time trajectory of the particle (right) for a fixed $R=2.5$ and $\tau=0.5$ (Type I - non-walking state), $\tau=2$ (Type II - steady walking state), $\tau=5$ (Type III - chaotic walk) and $\tau=8.9$ (Type IV - back-and-forth oscillations with a net drift). Time evolution for two different initial conditions $(x_d(0),X(0),Y(0),Z(0))=(0,1,0,0)$ (blue) and $=(0,-1,0,0)$ (red) is shown. In left panels, black filled circles represent stable equilibrium points whereas gray filled circles represent unstable equilibrium points of the Lorenz system.}
 \label{Fig: traj}
 \end{figure}
 
\textit{Active particle model.} A vertically vibrating liquid bath can support bouncing and walking droplets of the same liquid on its free surface~\citep{Couder2005,Couder2005WalkingDroplets, superwalker}. Each impact in the periodic bouncing of the droplet generates a localized standing wave that decays in time. The droplet interacts with these self-generated waves on subsequent bounces to propel itself horizontally resulting in walking and superwalking droplets~(see Fig.~\ref{Fig: schematic}(a)). Such droplets have been shown to mimic hydrodynamic analogs of several quantum systems~\citep{Bush2020review,Bush2024}. Three key features to note about this system - (i) the droplet and its underlying waves co-exist as a wave-particle entity; without the droplet, surface waves decay completely. (ii) There is memory in the system since at large amplitudes of bath vibrations, the droplet-generated waves decay very slowly in time. Hence the droplet motion is guided by the complex dynamical landscape sculpted by the history of waves along its trajectory. (iii) The droplet is self-driven or active~\citep{doi:10.1146/annurev-conmatphys-070909-104101} since the droplet locally extracts energy from the vibrating bath and converts it into persistent horizontal walking motion. A commonly used theoretical model that captures the key features of a droplet's walking dynamics is the stroboscopic model~\citep{Oza2013}. In this model, the vertical periodic bouncing motion of the droplet is averaged over and one gets a trajectory equation for the horizontal walking motion describing a particle driven by its wave-memory. As shown schematically in Fig.~\ref{Fig: schematic}(b), we consider a simple setup where the particle can move only in one horizontal dimension, resulting in the following trajectory equation of motion (in dimensionless units)~\citep{Oza2013,Valani_inf_mem}
\begin{equation}\label{eq: dim1}
    \ddot{{x}}_d + \dot{{x}}_d = F_{self} + F_{bias} ,
\end{equation}
where $x_d$ is the position of the particle and an overdot denotes a time derivative. To non-dimensionalize the variables, the spatial length $x$ is scaled by the inverse of the wavenumber $k_F=2\pi/\lambda_F$ with $\lambda_F$ as the wavelength of particle-generated waves, and time $t$ is scaled by $m/D$ where $m$ is the particle mass and $D$ is an effective drag coefficient averaged over the periodic bouncing motion~(see \citep{Oza2013} and \citep{Valani_inf_mem} for more details). The left-hand-side of Eq.~\eqref{eq: dim1} is composed of an inertial term $\ddot{{x}}_d$ and a dissipative force term proportional to particle velocity $\dot{{x}}_d$. The first term on the right-hand-side is the wave-memory force on the particle due to its self-generated waves. At each instant of time, the particle generates waves $W(x)$ centered at the particle location and decaying exponentially in time. By integrating over the history of the particle's trajectory, one can calculate the overall wave field generated by a linear superposition of the individual waves. The wave-memory force is then proportional to the gradient of this overall wave field at the particle location and can be written as
$$F_{self}=-R \int_{-\infty}^{t} W'\left( ({x}_d(t)-{x}_d(s)) \right)\,\text{e}^{-\frac{(t-s)}{{\tau}}}\,\text{d}s.$$
Here, the parameter $R$ is a dimensionless wave-amplitude of particle-generated waves and the parameter $\tau$ is a dimensionless wave decay rate and hence determines the memory of the system~(see \citep{Oza2013} and \citep{Valani_inf_mem} for more details on the parameters). The second term on the right hand side is the external bias force applied to the particle and we take it to be a constant force i.e. $F_{bias}=F$. 

\textit{Connection with Lorenz system.} In experiments, droplet-generated waves have features of both spatial oscillations and spatial decay~(see Fig.~\ref{Fig: schematic}(a)). However, if we only retain spatial oscillations and consider an idealized cosine waveform i.e. $W(x)=\cos(x)$, then this still captures the key qualitative features of walking dynamics~\citep{ValaniUnsteady} and significantly simplifies the system by allowing us to map the integro-differential equation in Eq.~\eqref{eq: dim1} into a system of nonlinear ordinary differential equations~\citep{phdthesismolacek,Durey2020lorenz,ValaniUnsteady,Valanilorenz2022}. This is done by differentiating with time, the wave-memory force $$Y=F_{self}=R \int_{-\infty}^{t} \sin(x_d(t)-x_d(s))\,\text{e}^{-\frac{(t-s)}{{\tau}}}\,\text{d}s,$$ and the wave field height at the particle location $$Z=R\int_{-\infty}^{t} \cos(x_d(t)-x_d(s))\,\text{e}^{-\frac{(t-s)}{{\tau}}}\,\text{d}s,$$ using the Leibniz integral rule giving the following system of ordinary differential equations~(see \citep{Valani_inf_mem} for a derivation):
\begin{equation}
    \label{lorenz}
    \begin{split}
    \dot{X}&=Y-X+F, \\
    \dot{Y}&=-\frac{1}{\tau}Y+XZ, \\
    \dot{Z}&=R-\frac{1}{\tau}Z-XY,
  \end{split}
\end{equation}
where $X=\dot{x}_d$ is the particle velocity. With no applied bias force, $F=0$, the system in Eq.~\eqref{lorenz} is equivalent to the celebrated Lorenz system~\citep{Lorenz1963}~(see Supplemental Material). Hence, the motion of our memory-driven active particle can be directly mapped onto the dynamics of the Lorenz system. We numerically solve the dynamical system of Eq.~\eqref{lorenz} in MATLAB using the inbuilt ode45 solver.

\begin{figure}
 \centering
 \includegraphics[width=\columnwidth]{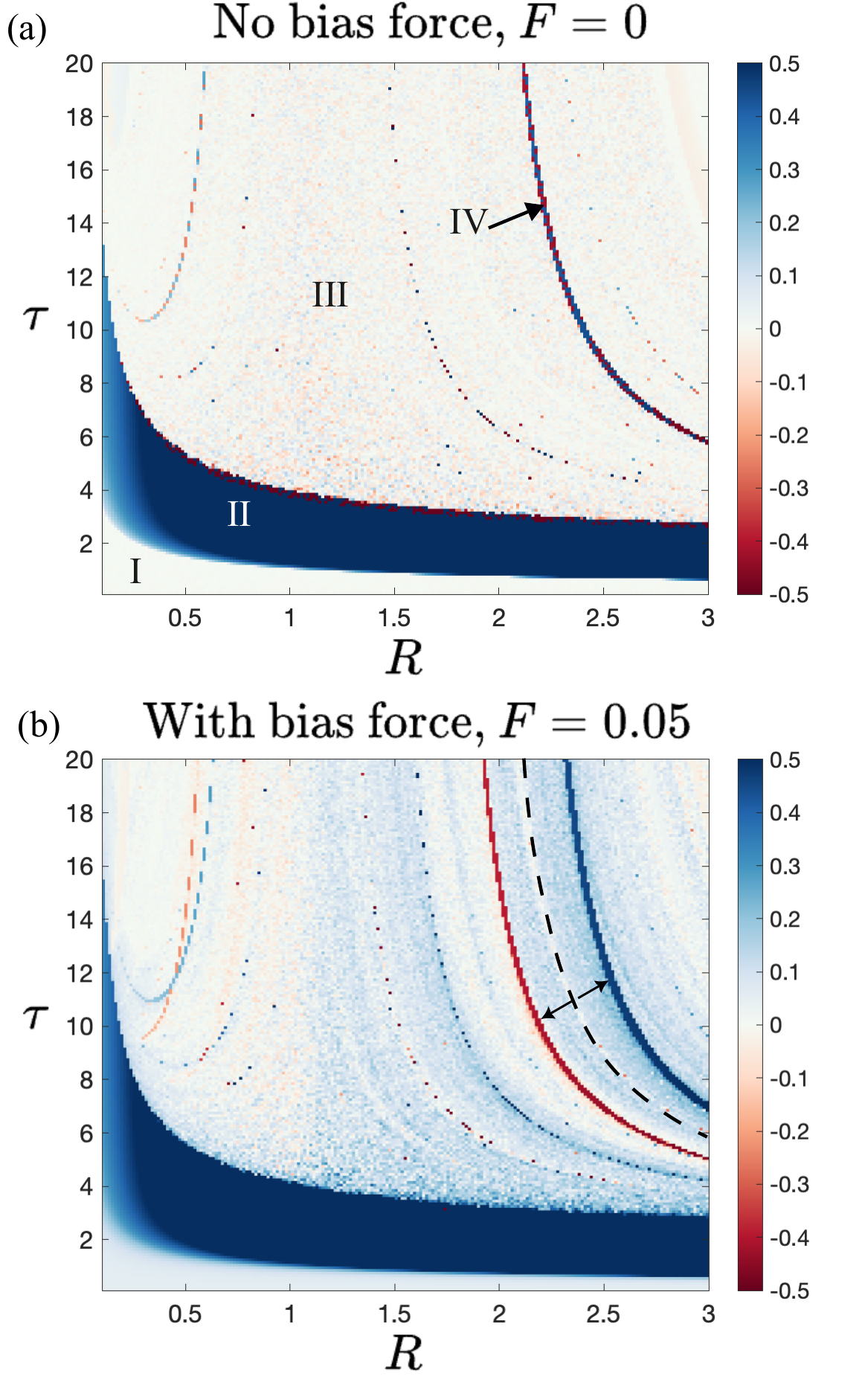}
\caption{Splitting of asymmetric limit cycles in the parameter space induce anomalous transport under applied bias force (see also Supplemental Video). Time-average particle velocity (colorbar) in the $(R,\tau)$ parameter space of the Lorenz model with (a) no bias force, $F=0$, and (b) a small applied bias force, $F=0.05$. The black dashed curve in (b) corresponds to a band of Type IV behavior in (a). These results are from simulations till $t=5000$ where only the latter half of the time series was used to calculate the average velocity of the particle. Initial conditions were fixed to $(x_d(0),X(0),Y(0),Z(0))=(0,1,0,0)$.}
 \label{Fig: PS}
 \end{figure}

With the above correspondence between the particle's trajectory equation and the Lorenz system, we can map the attractors and phase-space features of the Lorenz system onto the motion and trajectories of the active particle~\citep{ValaniUnsteady,Valani_inf_mem}. We proceed by reviewing different attractors realized in the Lorenz model~\citep{Sparrowbook,Valani_inf_mem} of Eq.~\eqref{lorenz} as a function of the memory parameter $\tau$ and map them onto the different types of trajectories for the particle. For small values of $\tau$, there is only one equilibrium point of the system at $(X,Y,Z)=(0,0,0)$ which is stable. This corresponds to a (Type I) non-walking state for the particle since its velocity is $\dot{x}_d=X=0$~(see Fig.~\ref{Fig: traj}(a)). With increasing $\tau$, the Lorenz system undergoes a pitchfork bifurcation (at $R=1/\tau^2$) where the non-walking state becomes unstable and a symmetric pair of stable equilibrium points emerge, $(X,Y,Z)=(\pm\sqrt{R-1/\tau^2},\pm\sqrt{R-1/\tau^2},1/\tau)$; this corresponds to steady walking motion (Type II) of the particle to right/left~(see Fig.~\ref{Fig: traj}(b)). Further increase in $\tau$ leads to a complex set of homoclinic, heteroclinic and Hopf bifurcations in the Lorenz system~\citep{Sparrowbook,jackson_1990} and eventually chaos emerges at large $\tau$ with a strange attractor in the phase space. Phase-space dynamics on the Lorenz chaotic attractor corresponds to a  (Type III) chaotic walk~\citep{ValaniUnsteady,Valaniattractormatter2023} for the active particle where it unpredictably switches between left and right moving states as the phase space trajectory jumps between left and right ``wing" of the Lorenz attractor~(see Fig.~\ref{Fig: traj}(c)). In experiments with a walking droplet moving in free space, the non-walking state, steady walking state and hints of a Hopf bifurcation leading to a walking state with velocity oscillations have been observed, whereas access to the chaotic regime and beyond is not yet achievable~\citep{Bacot2019, Hubert2019}. 

\begin{figure}
\centering
\includegraphics[width=\columnwidth]{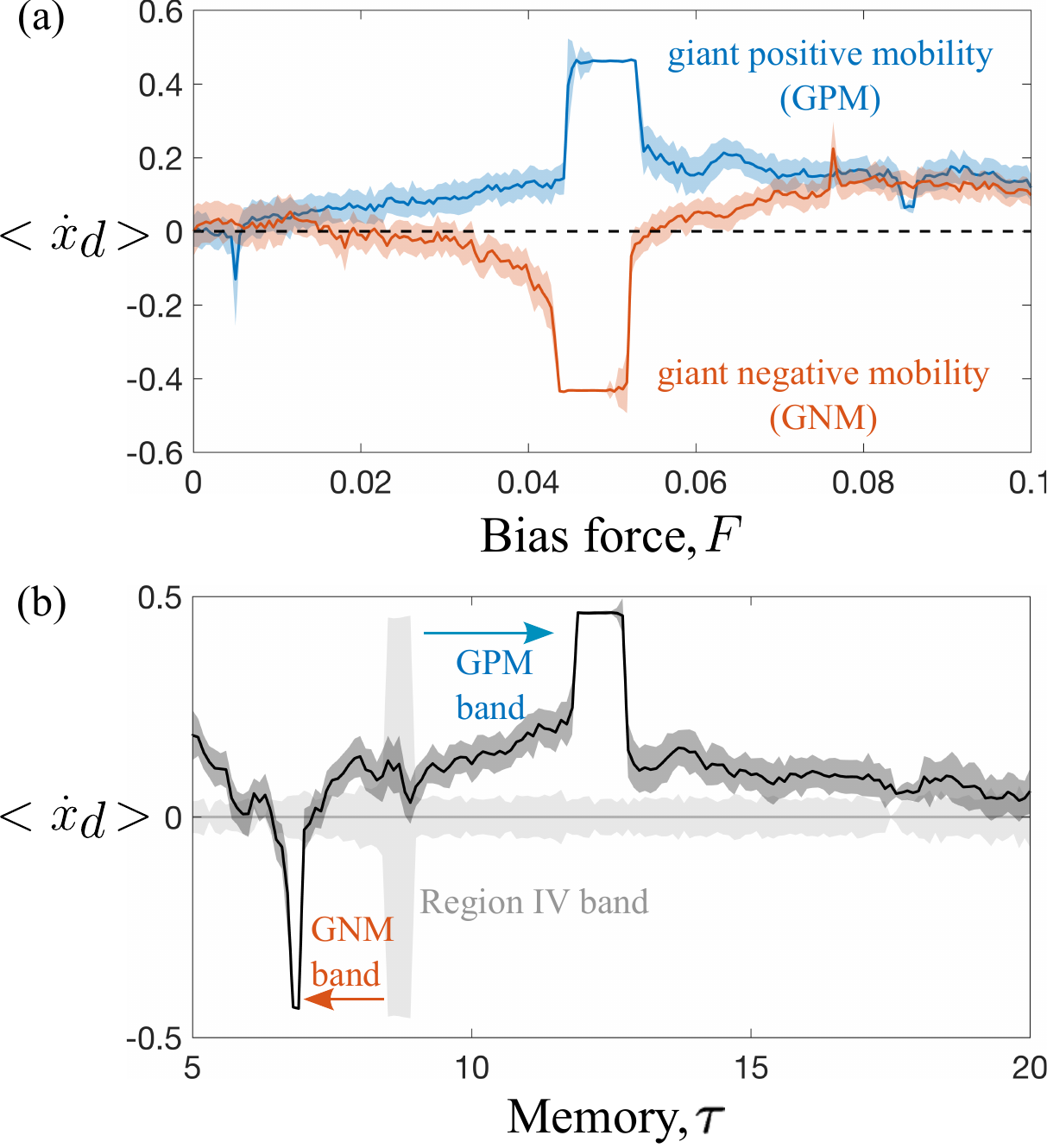}
\caption{Emergence of giant negative mobility (GNM) and giant positive mobility (GPM). (a) Average velocity (both time and initial-condition averaged) as a function of the bias force $F$ for fixed $R=2.5$ and, $\tau=6.9$ (red) and $\tau=12.2$ (blue). (b) Average velocity as a function of the memory parameter $\tau$ for fixed $R=2.5$ and, $F=0$ (gray) and $F=0.05$ (black). Shaded region indicates the spread in the average velocity for different initial conditions. The simulations were done till $t=2000$ and only the latter half of the velocity time series was used for time averaging. To average over initial conditions, $50$ different linearly-spaced initial conditions were chosen for $X(0)\in[-5,5]$ with fixed $x_d(0)=0$, $Y(0)=0$ and $Z(0)=0$.}
 \label{Fig: ANM GPM}
 \end{figure}

\textit{Dynamical mechanism for anomalous mobility.} It is common for periodic windows to exist within a chaotic regime of parameter space. Since the Lorenz system obeys a symmetry i.e. the system is symmetric about a rotation of $180$ degrees around the $Z$ axis, it restricts the types of periodic orbits that can exist. Two types of limit cycles can be found in the Lorenz system~\citep{Sparrowbook}: (i) a symmetric limit cycle about the $Z$ axis which would correspond to a back-and-forth oscillation for the particle with no net drift, and (ii) a symmetric pair of asymmetric limit cycles, as shown in Fig.~\ref{Fig: traj}(d), that corresponds to the particle undergoing back-and-forth oscillations along with a net drift either towards the left or right (Type IV). A parameter space plot showing the time-averaged particle velocity $\langle \dot{x}_d \rangle = \langle X \rangle$ in the $(R,\tau)$ space is shown in Fig.~\ref{Fig: PS}(a) which helps identify these oscillating states that have a net drift. Region I corresponds to non-walking state ($X=0$), Region II is steady walking ($X=\pm\sqrt{R-1/\tau^2}$), Region III is mostly filled with chaotic walks with no net drift ($\langle X \rangle = 0$). Of particular interest is Region IV bands corresponding to oscillating states with net particle drift. The mixture of positive (blue) and negative (red) drift states along this band indicates the sensitivity of these states to small changes in parameter values. 

Now under the application of a small bias force of $F=0.05$, the corresponding parameter space plot in Fig.~\ref{Fig: PS}(a) transforms to Fig.~\ref{Fig: PS}(b). We clearly see that the band corresponding to Region IV splits into two in the parameter space, one corresponding to the asymmetric limit cycle with a net drift opposite to the applied bias (red) and the other corresponding to the asymmetric limit cycle with a net drift in the direction of the applied bias (blue). Since, on these Region IV bands, the magnitude of net particle drift velocity is around $5$ times larger than neighboring parameter values, we identify the former as the giant negative mobility (GNM) band and the latter as the giant positive mobility (GPM) band. Further, on these separated bands, all initial conditions lead to the same limit cycle attractor, unlike the $F=0$ band where the basin of attraction of left/right drifting states in the initial-condition space are intricately mixed. After complete separation, the two bands migrate away from each other~(see Supplemental Video) with increasing small bias $F$ such that the GNM band moves below the $F=0$ band (black dashed curve) while the GPM band migrates above the $F=0$ band~\footnote{We note that with applied bias force and for small $\tau$, negative steady walking states (Type II) (i.e. in the direction opposite to the applied bias) exist as well. However, we don't classify these as anomalous behaviors since these states exist even for $F=0$, and furthermore, with increasing $F$, the magnitude of the negative steady velocity decreases for these states.}. 
We note that this band separation mechanism of coexisting asymmetric limit cycles is closely related to the separation mechanism of phase-locked periodic attractors reported by \citep{ANMnonchaos1} and \citep{ANMnonchaos2} for negative mobility of a passive particle in a periodic potential with external time-periodic driving. However, in contrast, the negative and positive mobility bands completely separate and migrate in the parameter space of our Lorenz model, whereas the two bands only partially separate in the system of \citep{ANMnonchaos1} along with the negative mobility band shrinking and positive mobility band growing in the parameter space. 

To demonstrate more clearly how this separation of bands manifests in GNM and GPM for our memory-driven active particle, let us consider the system at two different memory parameter $\tau$ values for a fixed $R$, one value below the region IV band with lower memory $\tau=6.9$ and the other above the band $\tau=12.2$ with higher memory. If we fixed the system parameter at these values, and investigate how the average velocity (both time and ensemble averaged) varies as a function of the bias force, we get the results shown in Fig.~\ref{Fig: ANM GPM}(a). For lower memory (red curve), the average velocity shows a strong dip of GNM near $F=0.05$. This is because, the GNM band after separation, migrates through the $(R,\tau)$ parameter space with increasing $F$ and passes through the $(R,\tau)=(2.5, 6.9)$ point in parameter space near $F=0.05$. Similarly, for the higher memory value (blue curve), the average velocity shows a strong spike near $F=0.05$ due to the passing of the GPM band. Figure~\ref{Fig: ANM GPM}(b) further shows the sensitivity of these behaviors as a function of the system memory $\tau$. With no bias force (gray curve), there is no net mobility of the particle (when ensemble averaged) at all values of memory $\tau$. However, with a small applied bias, the splitting of the region IV band results in GNM regions for lower memory and GPM regions for higher memory. This highlights the sensitivity of anomalous transport behaviors to the active particle's memory under small applied bias. 

\textit{Discussion and conclusions.} Although in our simplified Lorenz model we considered an idealized cosine wave, $W(x)=\cos(x)$, a wave form more closer to the experimental system of walking droplets that includes spatial decay shows a similar mechanism for GNM and GPM in a smaller region of the parameter space for weak spatial decay, whereas a strong spatial decay appears to suppress this behavior completely~(see Supplemental Material). Further, since the present dynamical mechanism lies within the chaotic regime of the parameter space, it would be difficult to observe this mechanism in experiments with walking/superwalking droplets moving in free space as access to the chaotic regime for free walkers/superwalkers is currently not achievable~\citep{Bacot2019}. 
However, it might be possible to experimentally demonstrate this effect in other active systems driven by low-dimensional chaos, such as autonomous mobile robots whose motion can be easily coupled to the Lorenz system~\citep{Valaniattractormatter2023,Paramanick2024,5430468,976022,Robot1}.

In summary, we have shown a dynamical mechanism rooted in the celebrated Lorenz system model for the generation of giant negative mobility and giant positive mobility for a memory-driven active particle under a small applied bias. {The degenerate pair of asymmetric limit cycles existing for the same parameter values completely separate and migrate in the parameter space upon small applied bias, resulting in the emergence of memory-dependent GNM and GPM for the active particle}. We believe this dynamical mechanism to be general to many low-dimensional chaotic systems and it might be at play in generating anomalous transport behaviors for active particles described by low-dimensional nonlinear systems. Further, such mechanisms may be used to tune the internal parameters of the system in order to control a particle's transport behaviors, or in sorting of active particles with different intrinsic properties~\citep{PhysRevLett.122.070602,PhysRevApplied.12.054002}.
\newline\newline
\textit{Acknowledgments}. This research was supported by the Leverhulme Trust [Grant No. LIP-2020-014]. 
Some of the numerical results were computed using the Hydra cluster in the Department of Physics at the University of Oxford.

\bibliography{apssamp}

\end{document}